# Inverse magnetocaloric effect in ferromagnetic Sm$_{0.6-x}$La$_x$Sr$_{0.4}$MnO$_3$ due to 4f-3d exchange interaction

V. B. Naik and R. Mahendiran[1]

Department of Physics, 2 Science Drive 3, Faculty of Science, National University of Singapore, Singapore 117542, Singapore

**Abstract**

We report magnetic and magnetocaloric properties of Sm$_{0.6-x}$La$_x$Sr$_{0.4}$MnO$_3$ (x = 0-0.6). A rapid increase around $T_C$ and an anomalous peak at a temperature $T^* \ll T_C$ occur in magnetization which lead to normal and inverse magnetocaloric effects (MCE), respectively. While $T_C$ increases with increasing x ($T_C$=118 K for x=0 and $T_C$=363 K for x=0.6), $T^*$ increases from 30 K (x=0) to 120 K (x=0.4) and then decreases to 105 K (x=0.5). The $\Delta S_m$ reaches +1.07 Jkg$^{-1}$K$^{-1}$ at 10 K and -4 Jkg$^{-1}$K$^{-1}$ around $T_C$ in x=0.4 for $\Delta H$=5T. The inverse MCE is attributed to antiferromagnetic coupling between Sm(4f$^5$) and Mn(3d$^{3+/4+}$) magnetic moments.

[1] E-mail: phyrm@nus.edu.sg

The search for energy efficient and environmental clean technology alternative to conventional vapour-based refrigeration has led to resurgence of interest in new technology called magnetic refrigeration (MR).[1] The MR is based on the concept of magnetocaloric effect (MCE) which is measured in terms of isothermal magnetic entropy change ($\Delta S_m$) and/or adiabatic temperature change ($\Delta T_{ad}$) in a magnetic material when it is exposed to a varying magnetic field ($H$). Large MCE is generally observed in materials, which undergo magneto-structural transition and/or first-order magnetic transitions, such as $Gd_5Si_{4-x}Ge_x$, $MnFeP_xAs_{1-x}$ etc.[1] Since MCE is maximum at the ferromagnetic (FM) Curie temperature ($T_C$), a stack of materials with varying $T_C$s, that is controlled by compositions, can be used for MR over a wide span of temperature ($T$ = 300-10 K). In this context, the hole-doped perovskite manganites, where the FM $T_C$ can be widely tuned by changing the one electron bandwidth, are considered to be emerging materials for MR technology.[2]

Majority of reports on the MCE in manganites are around paramagnetic-to-ferromagnetic (PM→FM) transition which leads to $\Delta S_m = S_m(H) - S_m(0)$ negative (decrease of magnetic entropy under $H$) *i.e.*, normal MCE (NMCE). In contrast to ferromagnets which cool upon demagnetization, antiferromagnets cools upon adiabatic magnetization *i.e.*, positive $\Delta S_m$ or inverse MCE (IMCE). Composites containing both NMCE and IMCE materials can be used to enhance refrigerant capacity (RC) as it is cooled by both adiabatic magnetization as well as demagnetization.[3] Generally, IMCE is observed in a compound which undergoes antiferromagnetic (AFM) transition directly from the paramagnetic (PM) state (ex., $Pr_{0.46}Sr_{0.54}MnO_3$)[4] or from FM state (ex., $Pr_{0.5}Sr_{0.5}MnO_3$)[5]. Very recently, IMCE was reported in $Ni_{50}Mn_{34}In_{16}$ alloy which undergoes austenite→martensite below the FM $T_C$.[3] The purpose of this letter is to reveal a possibility of observing IMCE in manganites at low temperature due to 4f-3d interaction between the magnetic moments of rare earth (*RE*) and transition metal ions within the long-rage FM ordered state. We have reported a huge MCE in the PM state due to unusual field-induced metamagnetic transition in $Sm_{0.6}Sr_{0.4}MnO_3$.[6] The $\Delta S_m$ was found to increase with decreasing size of the *RE* cation in $RE_{0.6}Sr_{0.4}MnO_3$ (*RE* = La, Pr. Nd, Sm etc) series.[7] Since the ionic radius of $La^{3+}$ ion (= 1.216 Å) is larger than that of $Sm^{3+}$ ion (= 1.132Å), the FM $T_C$ is expected to increase with varying x in $Sm_{0.6-x}La_xSr_{0.4}MnO_3$ providing the possibility of continuously tuning the $T_C$ and MCE over a wide temperature. In this letter, we report the occurrence of NMCE around $T_C$ and unusual IMCE at low temperature in $Sm_{0.6-x}La_xSr_{0.4}MnO_3$ with increasing size of the *RE* cation.

The polycrystalline compounds of $Sm_{0.6-x}La_xSr_{0.4}MnO_3$ (SLSMO) with compositions x = 0, 0.05, 0.1, 0.2, 0.3, 0.4, 0.5 and 0.6 were prepared by the solid state synthesis route and characterized by the standard X-day diffraction and ac magnetic susceptibility measurements. A commercial vibrating sample magnetometer (Quantum Design Inc., USA) was utilized for magnetization measurements. From the isothermal magnetic field dependence of magnetization, *M(H)*, measured at a temperature interval of $\Delta T$ = 5 K, the $\Delta S_m$ values are estimated using $\Delta S_m = \sum_i [M_i(T_i, H) - M_{i+1}(T_{i+1}, H)] \times \Delta H / (T_i - T_{i+1})$ where, $M_i$ and $M_{i+1}$ are the magnetization values measured at temperatures $T_i$ and $T_{i+1}$, respectively.

Fig. 1(a) shows the temperature dependence of magnetization, $M(T)$, measured while cooling and warming under $H = 1$ kG for x = 0, 0.05, 0.1, 0.2, 0.3, 0.4, 0.5 and 0.6 compounds. The rapid increase of $M(T)$ around $T_C = 118$ K for x = 0 is due to PM→FM transition and it exhibits hysteretic behavior while warming which indicates that the PM→FM transition is first-order in nature. Interestingly, $M(T)$ exhibits a peak around $T^* = 30$ K within the long-range FM ordered state. Such a peak was also observed in $Sm_{1-x}Sr_xMnO_3$ (x = 0.35-0.45) series around the same temperature by other researchers and they attributed it to increase in the coercivity.[8] The substitution of La dramatically increases the $T_C$ and the PM→FM transition becomes second-order for x ≥ 0.1 as indicated by the absence of hysteresis in $M(T)$. The $T_C$ determined from the minima of $dM/dT$ curves are 151 K, 185 K, 243 K, 288 K, 320 K, 345 K and 363 K for x = 0.05, 0.1, 0.2, 0.3, 0.4, 0.5 and 0.6, respectively. We have plotted $T_C$ and $T^*$ versus x in the inset on left and right scales, respectively. Surprisingly, the low temperature (low-$T$) peak $T^*$ also increases with x from $T^* = 30$ K for x = 0 to $T^* = 120$ K for x = 0.4 and then decreases to $T^* = 105$ K for x = 0.5, and finally it disappears for the end compound (x = 0.6).

Figs. 2(a)-(d) show isothermal $M(H)$ plots for four compositions, x = 0.05, 0.1, 0.3 and 0.4, respectively at selected temperatures. We have shown $M(H)$ data only at few selected temperatures in Fig. 2 for clarity. While the $M(H)$ of x = 0.05 varies linearly with $H$ above 200 K, a field-induced metamagnetic transition *i.e.*, a rapid increase of $M$ above a certain critical field, occurs in the temperature range of 150 K ≤ $T$ ≤ 200 K (above $T_C$). The metamagnetic transition is reversible upon decreasing $H$ with a small hysteresis of width 7 mT at $T = 160$ K and the hysteresis in $M(H)$ decreases with lowering $T$. While a similar but weak metamagnetic transition is seen in x = 0.1 (Fig. 2b) in the temperature range of 180 < $T$ < 225 K without any hysteresis, the metamagnetic transition is absent in x = 0.3 and 0.4. The $M(H)$ shows a FM behavior below $T_C$. However, we see that $M(H)$ at $T$ = 10 K < $T^*$ shows a cross-over behavior *i.e.*, the $M(H)$ curve at 10 K lies below the $M(H)$ curve at $T$ > 10 K in the field range of $\mu_0 H$ < 2 T.

Fig. 3 shows $\Delta S_m$ versus $T$ for $\Delta H$ = 5 T for x = 0, 0.05, 0.1, 0.3, 0.4 and 0.6 compounds. All the compounds show NMCE (-$\Delta S_m$ is positive) at their respective FM $T_C$s and IMCE (except x = 0.6) below their respective $T^*$s. A horizontal line at $\Delta S_m$ = 0 in Fig. 3 distinguishes the NMCE and IMCE. Since $M(T)$ of x = 0.6 compound did not show low-$T$ peak, we have not measured $M(H)$ curve below $T$ = 200 K. The peak value of $\Delta S_m$ at $T_C$ decreases from -6.5 Jkg$^{-1}$K$^{-1}$ for x = 0 to -4.7 Jkg$^{-1}$K$^{-1}$ for x = 0.1 and the $\Delta S_m$ spreads over a wider temperature with increasing x. The magnitude of $\Delta S_m$ at $T_C$ and $T$ = 10 K (below $T^*$) are plotted as a function of x in Fig. 3(a) on left and right scales, respectively. While the magnitude of $\Delta S_m$ at 10 K for x < 0.6 increases with x, the magnitude of $\Delta S_m$ at $T_C$ initially decreases with increasing x up to x = 0.3 and then increases slightly with further increase in x. The maximum IMCE is observed for x = 0.4 *i.e.*, $\Delta S_m$ = +1.07 Jkg$^{-1}$K$^{-1}$ at $T$ = 10 K for $\Delta H$ = 5. We have estimated RC using the equation $RC = \int_{T_2}^{T_1} \Delta S_m(T) dT$ where $T_1$ and $T_2$ are the temperatures corresponding to extremum values of half-maximum of the $\Delta S_m(T)$ peak around $T_C$ and it is shown for a field change of $\Delta H$ = 5 T as a function of x in Fig. 3(b). A significant value of $\Delta S_m$ = -4

Jkg$^{-1}$K$^{-1}$ at $T$ = 320 K along with a high RC of 214 Jkg$^{-1}$ for x = 0.4 makes it an interesting compound for room temperature MR.

The appearance of inverse MCE well below the long-range FM ordering is puzzling. Since a material can show IMCE if *dM*/*dT* is positive, antiferromagnets are expected to show the IMCE over a wide field range as shown recently in Pr$_{1-x}$Sr$_x$MnO$_3$ (x= 0.5 and 0.54) compounds.[4,5] However, IMCE in FM manganites is a rare phenomenon. Very recently, IMCE was reported in (La$_{0.7}$Sr$_{0.3}$MnO$_3$/SrRuO$_3$) superlattice due to a weak AFM coupling mediated by the spacer (SrRuO$_3$) layer between adjacent FM manganite layers.[9] The IMCE was also found at low temperatures in FM Pr$_{0.52}$Sr$_{0.48}$MnO$_3$ single crystal,[10] but this composition is close to AFM phase boundary in Pr$_{1-x}$Sr$_x$MnO$_3$ series (x = 0.5 is a layered *A*-type antiferromagnet) and a compositional fluctuations can induce AFM ordered local regions. We can rule of the possibility of AFM ordering in the Mn-sublattice because substitution of the large size La$^{3+}$ cation for Sm$^{3+}$ is expected to increase Mn-O-Mn bond angle and widen $e_g$-electron bandwidth. The rapid increase of $T_C$ with La-doping and monotonic decrease in low temperature resistivity with increasing x (not shown here) confirms that the FM double-exchange interaction among Mn-spins strengthens with increasing x. If the peak at *T\** was due to AFM ordering in the Mn-sublattice, *T\** should have decreased with increasing x, but this was not observed. The peak can neither be due to cluster nor spin glass transition since the sample has become magnetically homogenous with increasing x.

There are two possible origins for the IMCE observed in our samples. One possibility is that a spin-reorientation transition occurs in the Mn-spin lattice due to increase in magnetocrystalline anisotropy driven by $e_g$-orbital ordering and/or structural phase transition. The metallic ferromagnet $Pr_{0.5}Sr_{0.5}CoO_3$ also showed a magnetic anomaly at $T^* = 120$ K much below the FM transition ($T_C = 230$ K) and the rotation of magnetic moments from [110] to [100] axis around $T^*$ within the magnetic domain was observed by Lorentz microscopy.[11] Neutron diffraction studies on the same compound revealed changes in the magnetocrystalline anisotropy around $T^* = 120$ K driven by abrupt shortening of Pr-O bond length without change in symmetry[12] or by symmetry changing structural phase transition.[13] There is no available low temperature study on structural transition in $Sm_{0.6-x}La_xSr_{0.4}MnO_3$ series. Even if the spin-reorientation transition is caused by structural transition, the structural transition is most likely to be second-order for $0.1 \leq x \leq 0.5$ as suggested by the absence of hysteresis in $M(T)$. Another most likely possibility is that 4f-3d super-exchange interaction is operative in our compound. The decrease of $M(T)$ below $T^*$ can be understood as a result of antiparallel coupling of 3d spins of Mn sublattice and 4f-spins of Sm sublattice. Although these sublattices may order ferromagnetically, the coupling between them can be AFM. The ordering of $Sm(4f^5)$ moment should have been induced by the molecular field of Mn-sublattice. The ordering temperature increases with x and goes through a maximum for x = 0.4 and is absent in the La only compound (x= 0.6). There are some convincing evidences for 4f-3d interaction in manganites. Neutron diffraction studies by Suard *et al.*,[14] indicates that Nd moments order ferromagnetically below $T \approx 20$ K in $Nd_{0.7}Ba_{0.3-y}Sr_yMnO_3$, but the 4f-spins of Nd sublattice align antiparallel to the 3d-spins of Mn sublattice for y = 0 and it changes into parallel alignment for $y \geq 0.2$. Based on neutron diffraction and electron spin resonance line width studies, Dupont *et al.*,[15] suggested the ordering of Nd moments around 50 K in $Nd_{0.7}Ca_{0.3}MnO_3$. Neutron diffraction study by Cox et al.[16] indicated ordering of Pr moments around $T^* = 40$ K in AFM $Pr_{0.7}Ca_{0.3}MnO_3$. The AFM coupling between the Dy and Mn spins was suggested to be responsible for a drop of magnetization below 40 K in $Dy_{1-x}Sr_xMnO_3$ (x = 0.2).[17] However, the influence of RE moment ordering on magnetocaloric or electrical transport has not been reported so far. While the ordering of 4f moments is appeared to be influenced by the molecular field of Mn-sublattice, the exact mechanism of 4f-3d coupling is not clearly understood yet. A recent CGA+U calculation by Lin Zhu et al.,[18] emphasize on superexchange coupling between Nd-4f and Mn-3d ($t_{2g}$) electrons via O-2p electrons to explain FM coupling between 4f and 3d moments in $Nd_{0.67}Sr_{0.33}MnO_3$.

In summary, we have shown while the FM $T_C$ increases monotonically with increasing La content in $Sm_{0.6-x}La_xSr_{0.4}MnO_3$, an anomalous peak which appears around $T^* = 30$ K in x = 0 which initially shifts up with increasing x, reaches a maximum value of $T^* = 120$ K in x= 0.4 and then decreases. The $\Delta S_m$ is negative around $T_C$ and it decreases from $\Delta S_m = -6.2$ $Jkg^{-1}K^{-1}$ for x = 0 to -4.2 $Jkg^{-1}K^{-1}$ for x = 0.6 for $\Delta H = 5$ T. The inverse MCE occurs below $T^*$ and shows a maximum value of $\Delta S_m = +1.07$ $Jkg^{-1}K^{-1}$ at $T = 10$ K for $\Delta H = 5$ T in x = 0.4 which also shows a significant normal MCE ($\Delta S_m = -4$ $Jkg^{-1}K^{-1}$) at $T_C$. The inverse MCE has been suggested to be caused by the AFM coupling between 4f and 3d moments. The coexistence of normal MCE due to FM exchange-interaction between Mn spins and inverse MCE due to 4f-3d coupling in a single material is interesting since the sample can be cooled by adiabatic magnetization and demagnetization in different temperature regions which will enhance the refrigeration capacity.

**Acknowledgements:** RM acknowledges the office of DPRT (Grant no.: R144-000-197-123) for supporting this work.

**References:**

[1] K. A. Gschneidner, Jr., V. K. Pecharsky, and A. O. Tsokol, Rep. Prog. Phys. **68**, 1479 (2005), and references therein.

[2] M. H. Phan and S. C. Yu, J. Magn. Magn. Mater. **308**, 325 (2007).

[3] X. Moya, L. Mañosa, A. Planes, S. Aksoy, M. Acet, E. F. Wassermann and T. Krenke Phys. Rev. B **75**, 184412 (2007).

[4] V. B. Naik, S. K. Barik, R. Mahendiran and B. Raveau, Appl. Phys. Lett. **98**, 112506 (2011).

[5] N. S. Bingham, M. H. Phan, H. Srikanth, M. A. Torija and C. Leighton, J. Appl. Phys. **106**, 023909 (2009).

[6] A. Rebello and R. Mahendiran, Appl. Phys. Lett. **93**, 232501 (2008).

[7] H. Sakai, Y. Taguchi and Y. Tokura, J. Phys. Soc. Jpn. **78**, 113708 (2009).

[8] E. M. Levin and P. M. Shand, J. Magn. Magn. Mater. **311**, 675 (2007); R. P. Borges, F. Ott, R. M. Thomas, V. Skumryev and J. M. D. Coey, J. I. Arnaudas and L. Ranno, Phys. Rev. B **60**, 12847 (1999).

[9] S. Thota, Q. Zhang, F. Guillou, U. Luders, N. Barrier, W. Prellier, A. Wahl and P. Padhan, Appl. Phys. Lett. **97**, 112506 (2010).

[10] M. Patra, S. Majumdar, S. Giri, G. N. Iles and T. Chatterji, J. Appl. Phys. **107**, 076101 (2010).

[11] M. Uchida, R. Mahendiran, Y. Tomioka, Y. Matsui, K. Ishizuka and Y. Tokura, Appl. Phys. Lett. **86,** 131913 (2005).

[12] C. Leighton, D. D. Stauffer, Q. Huang, Y. Ren, S. El-Khatib, M. A. Torija, J. Wu, J. W. Lynn, L. Wang, N. A. Frey, H. Srikanth, J. E. Davies, Kai Liu, and J. F. Mitchell, Phys. Rev. B **79**, 214420 (2009).

[13] A. M. Balagurov, I. A. Bobrikov, D. V. Karpinsky, I. O. Troyanchuk, V. Yu. Pomjakushin and D. V. Sheptyakov, JETP Letters. **88,** 531 (2008).

[14] E. Suard, F. Fauth, C. Martin, A. Maignan, F. Millange, L. Keller, J. Magn. Magn. Mater. **264**, 221 (2007).

[15] F. Dupont, F. Millange, S. de Brion, A. Janossy and G. Chouteau, Phys. Rev. B **64**, 220403 (2001).

[16] D. E. Cox, P. G. Radaelli, M. Marezio and S-W. Cheong, Phys. Rev. B **57**, 3305 (1998).

[17] S Rößler, S Harikrishnan, U. K Rößler, Suja Elizabeth, H L Bhat, F Steglich and S Wirth, J. Phys.: Conf. Series **200**, 012168 (2010).

[18] L. Zhu, L. Li, T. Cheng and G. Wei, Phys. Lett. A **374**, 2972 (2010).

**Figure captions:**

Fig. 1 (Color online) The field-cooled magnetization $M(T)$ plots of $Sm_{0.6-x}La_xSr_{0.4}MnO_3$ compounds (x = 0 to 0.6) under $H$ = 1 kG. The inset shows the $T_C$ and $T^*$ as a function of composition x.

Fig. 2 (Color online) The $M(H)$ plots at selected temperatures for (a) x = 0.05, (b) x = 0.1, (c) x = 0.3 and (d) x = 0.4 compounds.

Fig. 3 (Color online) Temperature dependence of the magnetic entropy ($\Delta S_m$) obtained from $M(H)$ data at $\Delta H$ = 5 T for x = 0, 0.05, 0.1, 0.3, 0.4 and 0.6. The inset shows (a) values of the refrigeration capacity (RC), and (b) $\Delta S_m$ at $T_C$ (left scale) and $T$ = 10 K (right scale) as a function of composition x.

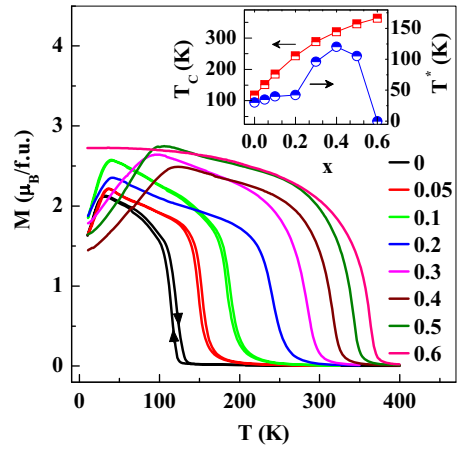

**Fig. 1**
**V. B. Naik** *et al.*

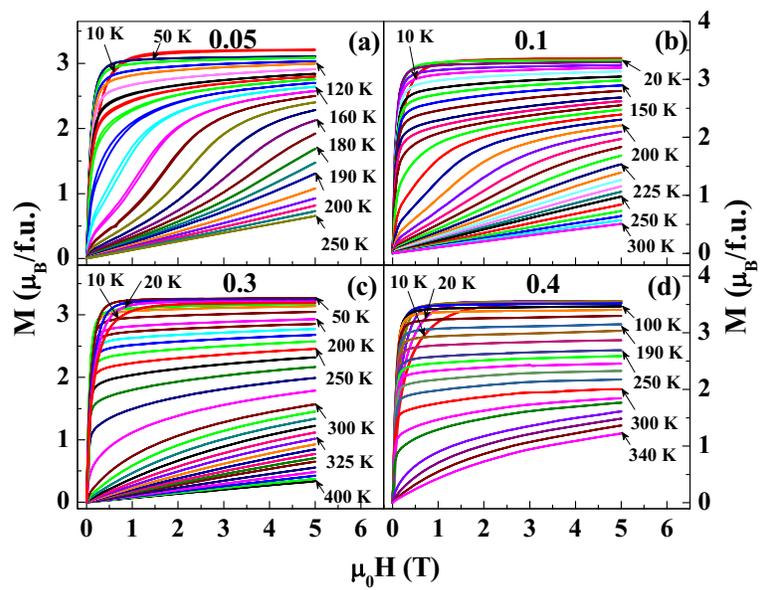

Fig. 2
V. B. Naik *et al.*

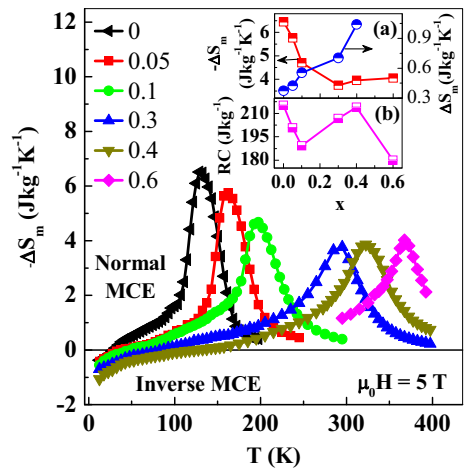

**Fig. 3**
**V. B. Naik** *et al.*